\documentclass[aps,12pt,onecolumn,preprintnumbers,amsmath,amssymb,superscriptaddress
]{revtex4}

\usepackage{graphicx}
\usepackage{ragged2e}
\usepackage{xcolor}
\usepackage{times}
\usepackage{bbold}
\usepackage{hyperref}
\usepackage{xr-hyper}

\makeatletter
\newcommand*{\addFileDependency}[1]{
  \typeout{(#1)}
  \@addtofilelist{#1}
  \IfFileExists{#1}{}{\typeout{No file #1.}}
}
\makeatother

\newcommand*{\myexternaldocument}[1]{
    \externaldocument{#1}
    \addFileDependency{#1.tex}
    \addFileDependency{#1.aux}
}

\myexternaldocument{SM}

\DeclareUnicodeCharacter{2212}{-}

\newcommand{\Imag}{\operatorname{Im}}

\begin{document}

%Title
\title{Super-resolution Control of Individual Two-dimensional Quantum Emitters}

%Author List/Affiliations
\author{Bosai Lyu$^{\dagger}$}
\affiliation{Department of Quantum Matter Physics, University of Geneva, Geneva, Switzerland}
%\affiliation{Department of Physics, Emory University, 30322 Atlanta, Georgia, USA}

\author{Valeria Vento$^{\dagger}$}
\affiliation{Department of Quantum Matter Physics, University of Geneva, Geneva, Switzerland}

\author{Ludivine Fausten}
\affiliation{Department of Quantum Matter Physics, University of Geneva, Geneva, Switzerland}

\author{Daniel Suárez-Forero}
\affiliation{Department of Quantum Matter Physics, University of Geneva, Geneva, Switzerland}

\author{Klevis Domi}
\affiliation{Department of Quantum Matter Physics, University of Geneva, Geneva, Switzerland}

\author{Kenji Watanabe}
\affiliation{Research Center for Electronic and Optical Materials, National Institute for Materials Science, 1-1 Namiki, Tsukuba 305-0044, Japan}

\author{Takashi Taniguchi}
\affiliation{International Center for Materials Nanoarchitectonics, National Institute for Materials Science,  1-1 Namiki, Tsukuba 305-0044, Japan}

\author{Alberto Morpurgo}
\affiliation{Department of Quantum Matter Physics, University of Geneva, Geneva, Switzerland}
\affiliation{Geneva Quantum Centre, University of Geneva, Geneva, Switzerland}

\author{Iaroslav Gaponenko}
\affiliation{Department of Quantum Matter Physics, University of Geneva, Geneva, Switzerland}

\author{Patrycja Paruch}
\affiliation{Department of Quantum Matter Physics, University of Geneva, Geneva, Switzerland}
\affiliation{Geneva Quantum Centre, University of Geneva, Geneva, Switzerland}

\author{Ajit Srivastava$^{\ast}$}
\affiliation{Department of Quantum Matter Physics, University of Geneva, Geneva, Switzerland}
\affiliation{Geneva Quantum Centre, University of Geneva, Geneva, Switzerland}

\maketitle
\justify
$^{\dagger}$These authors contributed equally to this work \\
$^{\ast}$Correspondence to: ajit.srivastava@unige.ch \\

% Abstract:

%TC:ignore
% ^ this is for the word counter

%{\bf 
%In transition-metal dichalcogenide (TMD) heterobilayers, lattice mismatch and reconstruction create localized traps for interlayer excitons, forming quantum emitters with potential as nanoscale sensors of correlated phases of matter. These emitters are electrically tunable and inherit spin–valley selection rules and strong light–matter coupling, yet deterministic spatial control remains challenging due to subwavelength confinement and disorder. Here, we present a platform that combines cryogenic optical spectroscopy with scanning probe microscopy to investigate trapped interlayer excitons in WSe$_2$/MoSe$_2$ bilayers. By exploiting AFM-based local Stark shift, we achieve super-resolution localization of emitters separated by only a few tens of nanometers and demonstrate deterministic control of individual charge states, including trion formation. Time-resolved measurements further allow us to resolve radiative and non-radiative decay channels, revealing controlled manipulation of individual exciton dynamics and opening a path toward coherent inter-dot coupling. These results establish a platform for sensing and trapping anyons in semiconducting FCIs.
%}

{\bf Localized interlayer excitons in semiconducting transition-metal dichalcogenide heterobilayers are quantum emitters with a static electric dipole moment, making them excellent nanoscale charge sensors to probe correlated quantum phases in a proximal layer. These emitters are electrically tunable and inherit spin–valley selection rules, yet their deterministic spatial control remains challenging due to subwavelength confinement. Here, we present a platform that combines cryogenic optical spectroscopy with scanning probe microscopy to investigate trapped interlayer excitons in MoSe$_2$/WSe$_2$ bilayers. By exploiting tip-induced local Stark shift, we achieve super-resolution of emitters separated by a few tens of nanometers and demonstrate deterministic control of individual charge states, including trion formation. Time-resolved measurements reveal tip-induced modification of the electromagnetic vacuum around individual emitters, and thus control of their radiative emission. Our multi-point charge sensing platform with optical readout is particularly well-suited for future study of fractionalization and anyon dynamics in semiconducting fractional Chern insulators. 
}

%TC:endignore
% ^ this is for the word counter
%opening a path towards coherent inter-dot coupling.

% Main Text:
\newpage
Two-dimensional (2D) quantum materials host a growing variety of correlated and topological phases, including (generalized) Wigner crystals~\cite{regan_mott_2020, xu_correlated_2020, smolenski_signatures_2021, li_local_2021, huang_correlated_2021, li_imaging_2021, wang_correlated_2020, zhou_bilayer_2021, tsui_direct_2024}, chiral superconductivity~\cite{han_signatures_2025, xu_signatures_2026}, and fractional Chern insulators~\cite{cai_signatures_2023, zeng_thermodynamic_2023, xu_observation_2023, lu_fractional_2024}. Although these phases have been investigated using a range of techniques, including transport measurements, scanning tunneling microscopy, and far-field optical spectroscopy, novel approaches are needed to probe their dynamics at the nanoscale. Among these approaches, exciton spectroscopy has played an important role in the discovery and characterization of several of these quantum states and can provide dynamical information through photon-correlation measurements~\cite{kass_many-body_2024, vandoolaeghe_photon_2026, kiper_photon_2026}. This non-invasive technique is particularly well-suited to studying quantum phases in semiconducting materials, such as transition metal dichalcogenides (TMDs), as it does not require high-quality electrical contacts. Moreover, optical excitations in TMDs can be used as sensors to probe proximal quantum materials~\cite{popert_optical_2022}.
%- The tip provides a ruler, that gives a distance information, so that we can say "this is correlation at this length scale".
%- dynamic information

However, the spatial resolution of far-field optical spectroscopy is fundamentally limited by diffraction, preventing the direct investigation of local fluctuations in charge and spin order at the nanoscale. Localized dipolar excitons, particularly those exhibiting spectrally narrow and temporally stable emission, are highly sensitive to their local electrostatic environment and can therefore serve as nanoscale probes of local charge dynamics~\cite{li_local_2021, seyler_signatures_2019}. These localized emitters typically form a spectrally distributed ensemble of sharp emission lines, enabling multiple emitters to be individually monitored with far-field optical spectroscopy, while simultaneously probing distinct spatial regions of the underlying quantum material.
Access to their spatial position and individual dynamics would further enable the measurement of multi-point spatiotemporal correlations. Such measurements could potentially provide a route to directly probe exotic quantum phases, including long-range entangled states such as fractional Chern insulators.

To this end, we implemented a novel experimental platform, dubbed Quantum Optical Super-resolution Microscope (QOSMic), which combines cryogenic far-field optical spectroscopy with scanning probe microscopy. A schematic of QOSMic is presented in Fig.~\ref{fig1}a (see Supplementary Fig. 1
%~\ref{figs:setup}
for details). QOSMic is a reflection confocal microscope for quantum emitters, with a conducting atomic force microscope (AFM) tip as a scanning local gate. 
The 2D sample hosting the emitters is encapsulated within thin insulating hBN layers preventing charge transfer while keeping a high spatial resolution of the tip. The emitters are illuminated through the transparent substrate and few-layer graphene gate using a laser beam focused through a high-numerical aperture objective. On the sample facing side, a self-sensing cantilever with a metallic AFM tip (see Methods) acts as a local top gate, while simultaneously modifying the local photonic density of states (LPDOS) through cavity effect. 
It is precisely the spatially localized and tunable nature of the tip-induced perturbation that allows us to achieve super-resolution control. 
For example, the metallic tip can generate a localized displacement field ${\bf E}({\bf r})$ through the sample, which depends on the distance ${\bf r}$ from the tip with both out-of- and in-plane components.
On the contrary, a dual-gated configuration, where top and bottom gates fully cover the sample, as in parallel-plate capacitor geometry, can only provide a global out-of-plane displacement field.
A quantum emitter at the position ${\bf r}$ couples to such an external displacement field through its permanent dipole moment ${\bf p}$. The result is a DC Stark shift of the exciton emission energy, which depends on the tip-exciton distance as 
\begin{equation}
    \mathcal{E} ({\bf r}) = - {\bf p}\cdot{\bf E({\bf r})}. \label{eq:shift}
\end{equation}
In addition to providing local spectroscopic information on the scale of tens of nanometers, well below the diffraction limit, QOSMic allows for a local control of the photoluminescence (PL) emission energy $\mathcal{E}$ and lifetime $\tau$ of individual emitters (Fig.~\ref{fig1}a).

%opening the possibility of studying exciton-exciton correlations with spatial resolution.

%This allows us to independently manipulate the steady-state population and the emission dynamics of excitons with nanoscale spatial resolution.

\begin{figure} 
\includegraphics[width = 6.5in]{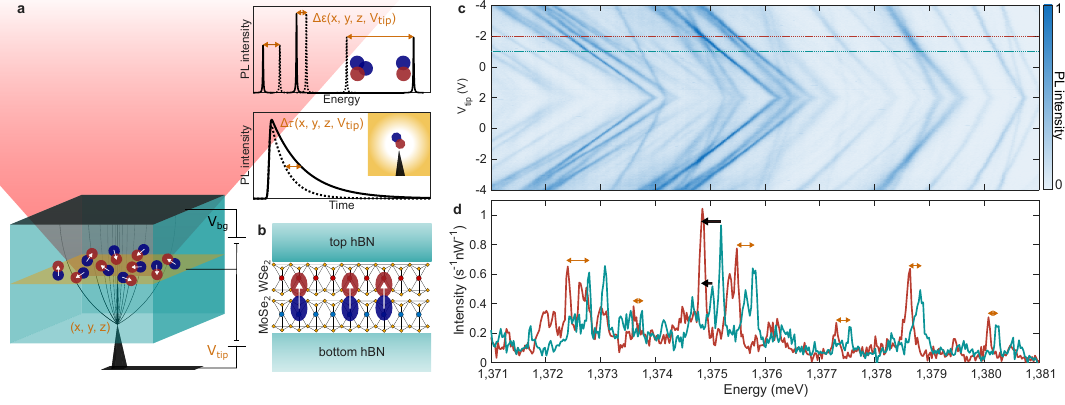} 
\caption{{\bf Quantum optical super-resolution microscope.} 
{\bf a,} Schematic of QOSMic. A metallic tip produces a non-homogeneous displacement field (black lines) through the sample. The sharp PL peaks of different quantum emitters show different energy shifts $\Delta \mathcal{E}$ due to Stark effect or doping when tuning tip position $(x,~y,~z)$ and voltage $V_\text{tip}$. At $z=0$, the tip is in contact with the sample. The variation of the lifetime $\Delta\tau$ is due to the modification of the local photonic density of states (LPDOS) near the tip (inset).
{\bf b,} Illustration of the MoSe$_2$/WSe$_2$ interlayer excitons with static electric dipole moment.
{\bf c,} Reproducible blue (red) shift of the PL emission of localized interlayer excitons in spot I of device A with increasing (decreasing) tip voltage $V_\text{tip}$, with the sample electrically floating and bottom gate grounded. Each spectral line corresponds to a different emitter in the optical spot.
{\bf d,} PL spectra at $V_\text{tip} = -2$~V (dark orange) and $-1$~V (green), corresponding to the line cuts in panel {\bf c}. The excitation energy and power are $1.698$~eV and $8$~nW respectively. The orange arrows indicate different magnitudes of the Stark shift for different dots. The black arrows show two dots tuned into resonance at $-2$~V.
%(a) we would like to show the schematic of the QOSMic. As a new type of microscope, it can achieve superresolution control of 2D quantum emitters. On top of this, it can do nonlocal quantum sensing that is not possible before. The sample should be more general, should not be limited to MoSe$_2$/WSe$_2$ heterostructure. But if the idea is rather complicated, maybe we should start putting data in Figure 2. To achieve nonlocal quantum sensing discussed in (a), the first step will be achieving superresolution localisation and control of quantum emitters, which will be demostrated in the following using a conductive AFM tip. (b and c) The idea was to show, with a conductive AFM tip as a local gate, that the Stark-shift of quantum emitters exhibits different slopes, in contrast to the global gate configuration. The emitter closer to the tip feels more electric field than the farther ones. Therefore, two dots can be turned in resonance which is ideal for the nonlocal quantum sensing.
} \label{fig1}
\end{figure}

{\bf Device geometry and characterization.}
To explore the advantages of QOSMic, we study hBN-encapsulated MoSe$_2$/WSe$_2$ heterobilayers, as shown in Fig.~\ref{fig1}b (see Methods and Supplementary Fig. 2
%~\ref{figs:sample_opt}
for details). %The crystallographic axes are aligned close to $0^\circ$ or $60^\circ$ angle, resulting in 
With a type-II band alignment, the lowest energy state for the electrons and the lowest energy state for the holes reside in different layers, respectively MoSe$_2$ and WSe$_2$. As a consequence, interlayer excitons in such heterstructures own a permanent out-of-plane static dipole moment ${\bf p}_z = e~d_z~\hat{z}$, where $d_z \simeq 0.7$~nm is the interlayer distance, which makes them sensitive to the out-of-plane component of an external displacement field. In order to obtain optically active interlayer excitons, especially trapped excitons (interlayer quantum emitters), the two monolayers are cryostallgraphically aligned \cite{seyler_signatures_2019, li_dipolar_2020, li_local_2021}. All samples measured are H-stacked, with a twist angle close to $60^\circ$, confirmed by their Landé g-factor of $\sim 16$ (see Supplementary Fig. 4).
%~\ref{figs:g-factor}

%In addition, close to $60^\circ$ twist angle, a permanent in-plane dipole moment has been predicted \cite{phillips_atomic_2026}, due to the atomic displacement caused by reconstruction into domains of aligned stacking \cite{zhao_excitons_2023}.  
%The thickness of the top and bottom hBN layers ($15$ to $20$~nm) is chosen to avoid charge transfer effects while keeping a high spatial resolution.  

%This type of heterostructure has been studied with global optical techniques \cite{seyler_signatures_2019, li_dipolar_2020, li_local_2021}. 
When exciting close to the resonance of the monolayer WSe$_2$ (around $1.7$~eV), at low enough excitation power (few tens of nW), the PL emission spectrum shows several narrow peaks near the free interlayer exciton energy at around $1.3$ to $1.4$~eV. These peaks have been shown to exhibit photon antibunching \cite{baek_highly_2020}, identifying them as individual quantum emitters or dots, associated with excitons localized in different traps within the optical spot.
Fig.~\ref{fig1}c, d show the PL spectrum of multiple quantum dots under the laser spot (spot I of device A). The emission linewidths are limited by the resolution of the spectrograph (see Methods). As the tip bias $V_\text{tip}$ is increased, the PL energy of each dot undergoes a linear Stark shift to the blue, with a distinct slope for each emitter. This slope is determined by the displacement field ${\bf E}$ locally experienced at the dot position. Since the magnitude of the electric field $\lvert{\bf E}\rvert$ decreases with increasing distance from the tip (see Supplementary Fig. 5),
%~\ref{figs:comsol}
a larger Stark-shift slope indicates a smaller tip–emitter separation. In contrast, for a device with global gate geometry the applied electrical field is homogeneous in the optical spot, therefore all quantum dots exhibit the same Stark-shift slope \cite{seyler_signatures_2019, li_dipolar_2020, li_local_2021, wang_moire_2021}. Here, by scanning the tip position and bias voltage, we can therefore independently tune the emission energies of individual dots and also bring them into resonance, as demonstrated in Fig.~\ref{fig1}d.

\begin{figure} 
\includegraphics[width = 6.5in]{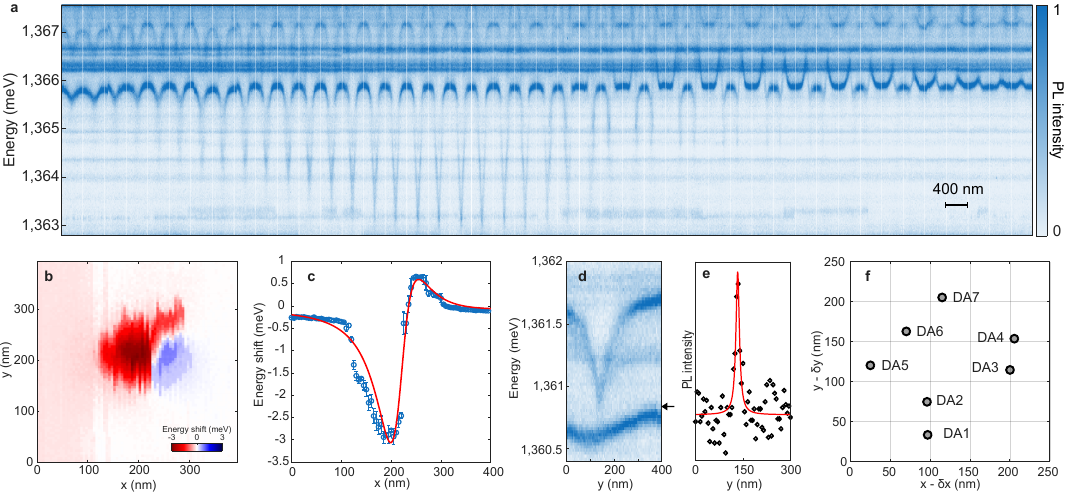}
\caption{{\bf Super-resolution of 2D quantum emitters.}
{\bf a,} Tip-position dependence of the PL emission spectrum from spot II of device A. The spectra are collected sequentially while the tip is raster scanned on a $400$~nm square, with $5$~nm step size and $V_\text{tip} = -1$~V. White lines separate the tip line scans along the $y$ direction (fast axis) at a fixed $x$ coordinate (slow axis). The panel displays only the central region of the raster scan, i.e. $225$~nm of the slow axis range.
{\bf b,} Tip-position-dependent energy shift of the dot at around $1,366$~meV extracted from panel {\bf a}. The backlash of the fast $y$ axis is corrected by shifting 3 pixels. 
{\bf c,} Line-cut (blue circles) of {\bf b} taken along the $x$ direction, averaged from $y = 200$ to $220$~nm. Error bars are the standard deviation of the average. The red curve is calculated from equation \ref{eq:shift} with a global scaling factor $A=0.22$, and electron-hole displacement ${\bf d} = (3.5,~ 0,~ 0.7)$~nm, while the electric field at the dot position is simulated as described in the Methods.
{\bf d,} Representative tip line-scan along the $y$ direction at $V_\text{tip}= -1.2$V, at spot III of device A, showing the Stark shift of dot DA1. 
{\bf e,} Line cut at fixed PL energy corresponding to the black arrow in {\bf d}. From Lorentzian fit (red curve), we can extract the coordinate $y_{DA1} \pm \Delta y_{DA1} = 132.9 \pm 0.7$.
%{\bf e,} Line cuts at fixed PL energies corresponding to the black arrows in {\bf d}. The FWHM of the most red-shifted peak is marked in red, and measure approximately $13$~nm.
{\bf f,} Extracted positions of seven emitters in a squared area of $250$~nm in spot III. The offsets on the axes are $\delta x = \delta y = 100$~nm. Error bars are smaller than the symbols. } \label{fig:superres}
\end{figure}

{\bf Super-resolution of 2D emitters.}
A first step toward exploiting these quantum emitters as local sensors of electronic states is to map their spatial locations and emission profiles. To this aim, we perform a raster scan of the tip along the $xy$ plane in contact with the hBN spacing layer ($z = 0$), while applying a negative bias voltage ($V_\text{tip} = -1$~V), and we measure the PL emission spectrum, as shown in Fig.~\ref{fig:superres}a. 
We first focus on the dot with emission energy of $1,365.8$~meV at the initial tip position, i.e. at a large tip-dot distance. The measurement reveals the Stark-energy shift of this dot as the tip approaches, with a maximum red shift of about $3$~meV. On the contrary, several other emitters under the laser spot exhibit negligible Stark shift, being far from the region of raster scan. 
Fig.~\ref{fig:superres}b shows the spatial profile of the emission energy shift of the dot as a function of the $(x,y)$ tip position. 
For a purely out-of-plane dipole ${\bf p} = {\bf p}_z$, we expect a red Stark-shift as the tip approaches the dot, such that the $(x,y)$ coordinates of maximum red shift identify the dot position. %In fact, a negative tip bias generates a ${\bf E}_z$ field parallel to ${\bf p}_z$, pulling the electron and hole apart. 

Remarkably, a blue-shift of about $0.8$~meV additionally appears, which can be attributed to the presence of an in-plane dipole component ${\bf p}_x = e~d_x~\hat{x}$ for a moir\'e-trapped exciton in H-stacked heterobilayer \cite{wang_intercell_2023, phillips_atomic_2026, rivera_valley-polarized_2016}.
%Since the in-plane field ${\bf E}_r$ is directed towards the tip, 
The measurement reveals that the in-plane dipole is oriented approximately along the $\hat{x}$ direction. As the tip approaches the dot while moving along $+(-)\hat{x}$, the dipole is parallel (antiparallel) to the in-plane electric field, resulting in an additional red (blue) Stark shift. 
%so that it's parallel (antiparallel) to the in-plane field when the tip approaches the dot from the left (right), causing an additional red (blue) Stark shift (see Supplementary Information).
We estimate an in-plane electron-hole displacement of $d_x \simeq 3.5$~nm (corresponding to a moir\'e length of approximately $6$ nm) by comparing the experimental data with the model given by equation \ref{eq:shift} (Fig.~\ref{fig:superres}c), where ${\bf p} = {\bf p}_x+{\bf p}_z$, and ${\bf E}$ is computed by finite element simulations for realistic tip geometry (see Methods). We remark that the observation of a different in-plane dipole orientation in spot III (see Supplementary Fig. 8)
%~\ref{figs:dipole_orientation}
suggests the possible role of local strain in determining the in-plane dipole orientation.

%rules out the possibility that the measured asymmetry in energy shift is due to a highly asymmetric in-plane field resulting from imperfections in the tip geometry.

To demonstrate the super-resolution capability of QOSMic, we consider a different region (spot III) of the same device, where in a single spectrum acquisition we can distinguish more than $78$ quantum emitters illuminated by the laser spot (see Supplementary Fig. 6).
%~\ref{figs:superres}
As an example, Fig.~\ref{fig:superres}d shows an increasing followed by a decreasing red Stark-shift of dot DA1, as the tip passes over it. By fitting the line cuts at fixed energy (Fig.~\ref{fig:superres}e), we extract the coordinates $x_{DA1},y_{DA1}$ and the uncertainties $\Delta x_{DA1},\Delta y_{DA1}$ of the narrowest PL profile, thus, obtaining the coordinates of maximum red Stark-shift as ($x_{DA1}\pm \Delta x_{DA1},y_{DA1}\pm \Delta y_{DA1}$), with $\Delta x_{DA1}, \Delta y_{DA1} \le 3$~nm. Note that the position of maximum Stark shift need not correspond to the position of the dot because of a finite in-plane dipole moment (Fig.~2c, Supplementary Fig. 5).
%~\ref{figs:comsol}
However, we find that the direction of the in-plane dipole does not vary much within a given spot (see Supplementary Fig. 8),
%~\ref{figs:dipole_orientation}
producing a constant offset in position, which does not affect the relative distance between the dots. 

%As the centroid of the maximum Stark shift can be located with high accuracy, 

%corresponding to maximum red Stark-shift.

%with $\Delta y_{DA1} \le 20$~nm

%sets our resolution limit. Note that the coordinate $y_{DA1} \pm \Delta y_{DA1}/2$ of maximum red shift identifies the position of dot $DA1$ up to an offset along the direction of the in-plane dipole. 
%Since the dipoles are oriented in the same directions within the same domain (see Supplementary Fig.~\ref{figs:dipole_orientation}), this offset is constant for all the dots under the laser spot. Therefore, since we are only interested in the relative distance between dots, we assume that $(x_{Di}, y_{Di})$ is the position of dot $Di$. 

%Therefore, 
We perform a raster scan of the tip on an area of $400 \times 400$-nm$^2$ within the laser spot, and extract the positions of $7$ different emitters by using the method described above (see Supplementary Fig. 7).
%~\ref{figs:superres_zoom}
These $7$ dots are located in a $250$-nm square at the center of the raster scan (Fig.~\ref{fig:superres}f). In particular, we can resolve the emitters DA1 and DA2 at a distance of $d_{DA1-DA2} \simeq (41.4 \pm 2.8)$~nm, and the emitters DA3 and DA4 at a distance of $d_{DA3-DA4} \simeq (39.5 \pm 1.8)$~nm, well below the limit imposed by diffraction (see Supplementary Note 1 for a discussion on spatial resolution). In addition to interlayer dots in MoSe$_2$/WSe$_2$ heterobilayers, QOSMic also allows the localization of dots in WSe$_2$ monolayers with or without a permanent dipole moment (see Supplementary Fig. 9
%~\ref{figs:monolayerdot}
and Supplementary Note 2 for details), demonstrating the versatility of the platform.

\begin{figure} 
\includegraphics[width = 6.5in] {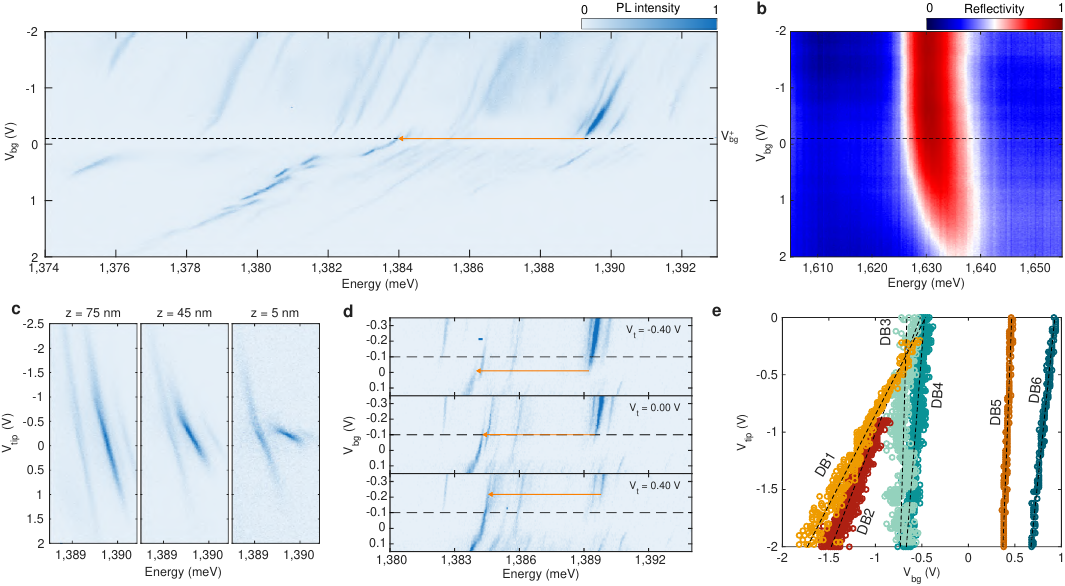}
\caption{{\bf Selective doping of 2D quantum emitters.}
{\bf a,} Bottom gate dependence of the PL emission from spot I of device B. The horizontal dashed line indicates the onset gate voltage $V_\text{bg}^+$ for global electron doping in MoSe$_2$. The orange arrow indicates the transition from neutral exciton to trion for one of the emitters at the onset voltage. %Most emitters exhibit similar change in emission energy and intensity. 
{\bf b,} Bottom gate dependence of the reflectivity of MoSe$_2$. Upon the increase in electron doping density, the neutral exciton resonance evolves into a dispersive repulsive polaron feature. The onset doping voltage is consistent with panel {\bf a}. 
{\bf c,} Tip voltage dependent PL emission of the two dots at about $1,390$~meV with the tip on axis with the higher-energy dot for different tip-sample distances $z$, at $V_\text{bg} = -0.5$~V. 
{\bf d,} Back voltage dependent PL emission with the tip in contact with the target dot, at three different $V_\text{tip}$ values.
{\bf e,} $\{V_\text{tip}, ~V_\text{bg}\}_i$ dependence of the emission energy of dot $i = B1- B6$. The dots $B1-B4$ are neutral excitons, while the dots $B5$ and $B6$ are trions. Dashed lines are linear fit, from which we can extract the slopes $\alpha_{Di}(x,y)$ (see main text for details).
%(a)We want to show using a single global bottom gate, electric field and doping always change simultaneously, instead of a dual-gate device E field and n can be separately tuned. 
%(b)To show the onset voltage of electron doping. Consistent with the doping onset in panel {\bf a}. 
%(c)
} \label{fig:doping}
\end{figure}

{\bf Selective doping.}
QOSMic allows us to locally dope the TMD sample and therefore,
the charge state of individual quantum dots. To enable doping control, we fabricate another device (device B) with bottom few-layer graphene gate and an additional contact to the TMD layer. Combining with the tip as a local top gate, a dual-gate configuration is achieved which allows separately to tune electric field and doping. In a typical dual-gated device with two global gates, the charge (doping) state is uniformly shared through the whole sample, and can be continuously tuned while keeping ${\bf E}=0$ by ramping the top and bottom gates in a symmetric fashion, i.e. $V_\text{tg}= \alpha V_\text{bg}$ where the constant $\alpha$ takes into account the difference in top and bottom hBN thicknesses.
%In a global gating configuration, the electric field and doping state are uniformly shared through the whole sample, and can be independently tuned by ramping the gates in an antisymmetric or symmetric fashion respectively, i.e. $V_\text{bg}=\mp \alpha V_\text{tg}$, where the constant $\alpha$ takes into account the difference in top and bottom hBN thicknesses. %and doping? 
In our local gating configuration, for a specific tip position $(x,y)$, each dot D$i$ at $(x_{Di}, y_{Di})$, a different $\alpha_{Di}(x,y)$ exists, such that simultaneously tuning $V_\text{tip}$ and $V_\text{bg}$, with $V_\text{tip}= \alpha_{Di}(x,y) V_\text{bg}$, defines the charge state of this dot.
%while the surrounding sample is subject to the variation of both doping and electric field. Note that equipotential lines around the tip position share the same slope $\alpha(x,y)$, so that for a perfectly symmetric tip shape $\alpha(x,y) \equiv \alpha(r)$.

Trion formation is employed here as the main indicator of doping. We begin by studying the global doping behaviour of the emitters with the tip withdrawn. By increasing $V_\text{bg}$ from -2V to $\simeq-0.1$~V, the dots emission energies (in spot I of device B) undergo a continuous Stark shift to the red (Fig.~\ref{fig:doping}a). In the meanwhile, the electron density increases globally in the sample. 
%The result of the mixup of electric field and doping on the PL emission is shown in Fig.~\ref{fig:doping}a, where we ramp the global bias $V_\text{bg}$ while the tip is withdrawn (or equivalently the tip is far away from the spot position). As $V_\text{bg}$ increases, the dots emission energies undergo a continuous Stark shift to the red. 
At $V_\text{bg}^{+}\simeq-0.1$~V, most emitters exhibit a discrete jump of approximately $5$~meV to the red, indicating the transition from charge neutrality to electron doping consistent with previous reports~\cite{li_local_2021, wang_moire_2021, jauregui_electrical_2019, liu_signatures_2021} (also see Supplementary Fig. 10).
%~\ref{figs:DCP}
This transition is further confirmed by the corresponding change in the reflectivity (Fig.~\ref{fig:doping}b), the degree of circular dichroism (DCP, see Supplementary Fig. 10),
%~\ref{figs:DCP}
and the lifetime (see Supplementary Fig. 11).
%~\ref{figs:All_dots_lifetime}
For example, the dot connected by the solid orange arrow is neutral exciton at approximately 1,389.5 meV and becomes positively charged trion at about 1,384 meV. Their energy difference of approximately 5.5 meV is the binding energy of this trion. %\ref{figs:dcp}Indeed, $5$~meV is roughly the energy required to bound an additional electron to a neutral exciton, forming a trion state. 

Next, we place the tip on top of this high intensity dot and scan $V_\text{tip}$ (Fig.~\ref{fig:doping}c), while approaching the tip along $z$ axis, at $V_\text{bg} = -0.5$ V. In this measurement, we focus on its charge neutrality plateau together with that of another dot located approximately 58 nm away (emitting at 1,389 meV). As expected, the Stark shift is in the opposite direction to that of $V_\text{bg}$, i.e, towards higher energy, when $V_\text{tip}$ is increased. However, the plateau of the target dot shrinks dramatically (approximately ten times) when the tip approaches, in stark contrast to the low energy dot. In other words, the tip selectively changes the onset voltage of electron and hole doping for a specific dot. 
To confirm this observation, we place the tip in contact (on top of the target dot), and repeat the $V_\text{bg}$ scan for different values of $V_\text{tip}$ (Fig.~\ref{fig:doping}d). 
For a negative (positive) $V_\text{tip}$, the hole doping of the target dot occurs at an onset voltage $V_\text{bg, target}^{+}$ larger (smaller) than $V_\text{bg}^{+}$. Therefore, by tuning $V_\text{tip}$, we can continuously shift $V_\text{bg, target}^{+}$ across the global threshold $V_\text{bg}^{+}$.

To better visualize the effect of the tip and its local modification of the electrostatic environment, we perform a PL measurement while simultaneously ramping $V_\text{tip}$ and $V_\text{bg}$.
Therefore, we track the values $\{V_\text{tip}, ~V_\text{bg}\}_{Di}$ needed to keep the emission of the $i$-th dot at a fixed energy. Fig.~\ref{fig:doping}e shows the extracted $\{V_\text{tip}, ~V_\text{bg}\}_{Di}$ for a few different dots, both in the neutral and trion states, at a fixed tip position $(x,y)$. Four of the dots ($DB3-DB6$) exhibit a very large slope $\alpha_{Di}(x,y)$ indicating that the tip is too far to affect them. 
On the other hand, two of the dots ($DB1, ~DB2$) are greatly affected by the tip. Therefore, by tuning $V_\text{tip}$ and $V_\text{bg}$ with the constant ratio $\alpha_{Di}(x,y)$, we can achieve selective doping of the $i$-th dot without introducing an additional electric field. 

\begin{figure}
\includegraphics[width = 6.5in]{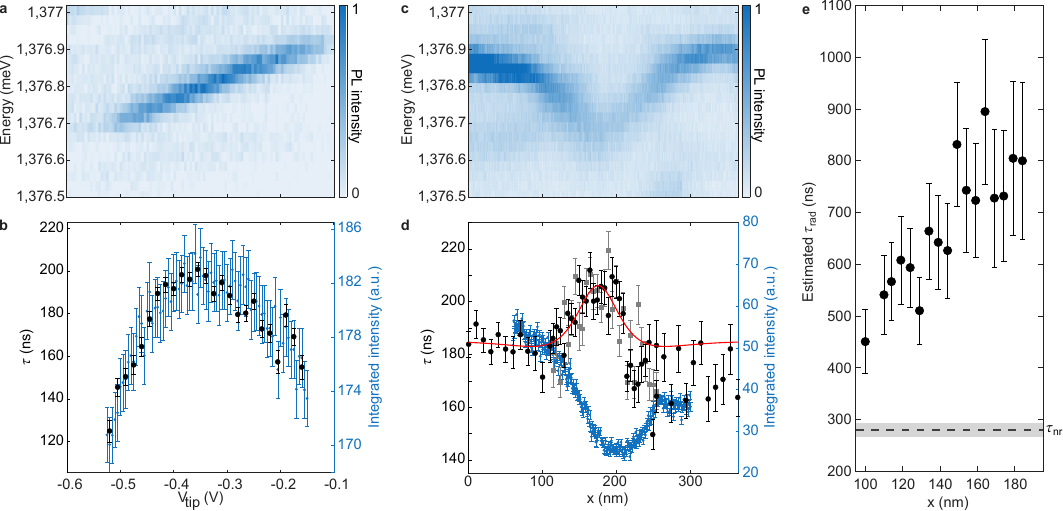}
\caption{{\bf Lifetime control of an individual quantum emitter.}
{\bf a,} PL emission spectra as a function of tip bias for a neutral dot in spot IV of device A. The tip-dot distance is approximately $200$~nm. 
{\bf b,} Total lifetime $\tau$ as a function of $V_\text{tip}$ (black circles), compared with the integrated PL intensity from panel {\bf a} (blue circles). The integration range spans $264.2~\mu$eV, and is centered at the peak of maximum intensity.
{\bf c,} PL emission spectra as a function of the tip $x$-coordinate, with the tip $y$-coordinate fixed close to the dot position. The tip voltage is fixed at $V_\text{tip}=-0.22$~V.
{\bf d,} Total lifetime as a function of increasing (black circles) and decreasing (gray squares) tip $x$-coordinate, compared with the integrated PL intensity from panel {\bf c} (blue circles). The red curve corresponds to the simulation described in Methods and Supplementary Fig. 13,
%~\ref{figs:lifeVSsim}
with a baseline $\tau_\text{far} = 185$~ns and $x$-axis offset of $175$~nm. In panels {\bf b} and {\bf d} error bars on $\tau$ are the uncertainties obtained from the exponential fits to the decay curves, whereas those on the integrated intensities represent one standard deviation from Monte Carlo analysis. 
{\bf e,} Radiative lifetime $\tau_\mathrm{rad}$ estimated from the measurements in panels {\bf c} and {\bf d} as a function of increasing tip $x$-coordinate. The dashed line indicates the estimated non-radiative lifetime, with uncertainty denoted by the gray shadow. 
This uncertainty and the error bars correspond to the error propagation from lifetime and integrated intensity. 
} \label{fig:lifetime}
\end{figure}

{\bf Control of radiative dynamics.}
The tunability of the LPDOS provides an additional control knob for the emission properties of quantum dots. Fig.~\ref{fig:lifetime}a shows the charge neutral plateau of a dot in spot IV of device A, while the tip is $200$-nm away. 
%The PL emission Stark shifts to the blue increases with $V_\text{tip}$, displaying a plateau of charge neutrality. This is interrupted by electron and hole doping at $V_\text{tip}\simeq -1.4$ and $V_\text{tip}\simeq -0.4$ respectively.
By selecting a narrow spectral region around the emission energy of the dot (see Methods for details), we measure the total lifetime $\tau$ as a function of $V_\text{tip}$ (Fig.~\ref{fig:lifetime}b). The lifetime shows the same qualitative behavior as the integrated PL intensity, dropping at the edges of the plateau. Indeed, when doping with either electrons (increasing $V_\text{tip}$) or holes (decreasing $V_\text{tip}$), the presence of extra free carriers opens extra non-radiative decay channels (also see Supplementary Fig. 11
%~\ref{figs:All_dots_lifetime}
for the $V_\text{bg}$ dependent lifetime)\cite{jauregui_electrical_2019}. These reduce both the emission intensity 
\begin{equation}
    I_\text{PL} \propto \frac{\Gamma_\text{rad}}{\Gamma_\text{rad}+\Gamma_\text{nr}} \label{eq:IPL}
\end{equation}
and the total decay rate 
\begin{equation}
\Gamma \equiv  \tau^{-1} = \Gamma_\text{rad}+\Gamma_\text{nr}, \label{eq:tau}
\end{equation}
where $\Gamma_\text{rad}=\tau_\text{rad}^{-1}$ and $\Gamma_\text{nr}=\tau_\text{nr}^{-1}$ are the radiative and non-radiative rates respectively. 

We select a value of $V_\text{tip}$  such that when we move the tip close to the dot, it emits at the center of the plateau, i.e. with constant non-radiative rate. The PL spectrum in Fig.~\ref{fig:lifetime}c exhibits a red Stark-shift when the tip approaches the dot, as expected since $-{\bf E} \cdot {\bf p}<0$.
At the same time, the PL intensity drops by $\simeq$ $2$ times, and the total lifetime (Fig.~\ref{fig:lifetime}d) increases by $12\%$. 
Since the non-radiative rate is constant, knowing $\tau$ and $I_\text{PL}$ allows us to estimate both $\tau_\text{nr}$ and $\tau_\text{rad}$ as a function of the tip position, using equations \ref{eq:IPL} and \ref{eq:tau}. Fig.~\ref{fig:lifetime}f shows a dramatic increase of the estimated radiative lifetime by almost a factor of $2$. Note that while the reduction in intensity can arise from a decrease in non-radiative lifetime, we observe an increase in total lifetime, further ruling out modifications to non-radiative effects.

To explain this increase in radiative lifetime, we can invoke the fact that the ${\bf E}$ which induces the red Stark-shift, pulls the exciton apart and reduces the electron-hole overlap, and consequently the oscillator strength (see Supplementary Fig. 12a). 
%~\ref{figs:lifeVSshift}
Indeed, such an effect has been observed in a dual-globally-gated MoSe$_2$/WSe$_2$ heterobilayer sample \cite{jauregui_electrical_2019}, however, the expected lifetime increase from such an effect, corresponding to our observed red-shift, is approximately 30 times smaller (see also Supplementary Fig. 12b-c),
%~\ref{figs:lifeVSshift}
suggesting another mechanism at work. Indeed, the radiative lifetime can also be affected by the modification of LPDOS around the emitter caused by the metallic tip. To quantify this contribution we perform numerical simulations based on dyadic Green's functions \cite{novotny_single_1996, novotny_principles_2012} (see Methods). As shown in Fig.~\ref{fig:lifetime}d, the computed change in total lifetime is in good agreement with the experimental data. Thus, we conclude that while the red-shift is caused by V$_\mathrm{tip}$, the change in radiative rate is primarily due to modification of LPDOS induced by the photonic cavity formed by the metallic tip in a dielectric environment.

{\bf Discussion and Outlook.} As QOSMic does not require quantum emitters with static dipole moment, it
should be extendable to other quantum emitters such as  spin defects in hBN~\cite{gottscholl_spin_2021}, thus giving access to dynamical multi-point spatiotemporal correlations of spins, in addition to charge.
Our preliminary demonstration of bringing two emitters into resonance should enable investigation of the interplay of static dipolar interactions and vacuum-mediated strong coherent dipole-dipole interactions in the deep sub-wavelength regime to probe collective emissive behavior such as superradiance~\cite{hettich_nanometer_2002,Kumlin_Superradiance_2025}. Such dipole-dipole interactions can be exploited for quantum coherent sensing of quantum matter beyond Stark-shift. By including other AFM modes such as piezoresponse force microscopy, electrostatic force microscopy, and conductive AFM in QOSMic to image moir\'e lattice in situ, valuable information correlating moir\'e structure with optical properties can be gained. Our scanning gate technique is particularly attractive to study the dynamics of anyons by nucleating and trapping them in optically active fractional Chern insulators in t-MoTe$_2$~\cite{li_signatures_2026,Mostaan_anyon_2026,Wagner_sensing_2026}. 
\justify
{\bf Acknowledgments}
We thank A.~Imamo\u{g}lu, T.~Smole\'nski, M.~Kroner, A.~Bergschneider and G-M.~Schn\"uriger for insightful discussions. A.~S. acknowledges funding from the State Secretariat for Education, Research and Innovation (SERI)-funded European Research Council Consolidator Grant TuneInt2Quantum (No.~101043957), Swiss National Science Foundation grant 200021-236651.  K.~W.~and T.~T.~acknowledge support from the JSPS KAKENHI (Grant Numbers 21H05233 and 23H02052) , the CREST (JPMJCR24A5), JST and World Premier International Research Center Initiative (WPI), MEXT, Japan.

% Author Contributions Statement
\justify
{\bf Author Contributions Statement} B.~L., V.~V. and A.~S. conceived the project, analyzed the data, and wrote the manuscript. K.~W., T.~T. provided the hBN crystals. B.~L. design the home-built insert with the assistance of V.~V., L.~F., I.~G., P.~P., and A.~S.. B.~L. and L.~F. prepared the samples. B.~L. and V.~V. built the setup and carried out the measurements with the assistance of K.~D. and D.~S.. V.~V. carried out the simulations. A.M. provided the fabrication facility and participated in scientific discussions.

% Competing Interests Statement
%\justify
%{\bf Competing Interests Statement}
%The authors declare no competing interests.

%\justify
%{\bf Data Availability}
%All the data that support the findings of this study are reported in the main text and Supplementary information. Source data are provided with this paper. 

\bibliographystyle{apsrev}
\bibliography{references}

\justify
\textbf{Methods}
\\
\textbf{Quantum optical super-resolution microscope} \\
The cryogenic QOSMic consists of two nanopositioner towers with three translational degrees of freedom (XYZ) (Supplementary Fig. 1)
%~\ref{figs:setup}
One tower carries a sample on a transparent sapphire substrate. The other tower holds an AFM cantilever. The sample faces down towards the AFM tip. On the other side of sapphire, an achromatic objective (NA = $0.81$, WD = $0.7$~mm) is equipped for sample and tip observations and optical measurements. The entire assembly is cooled down to 4 K in a high-purity helium environment. The conductive AFM tip ($100$~nm Pt) remains in continuous contact with the hBN surface throughout all line-and raster tip scan experiment. Self-sensing AFM cantilevers equipped with piezoresistive elements are used for monitoring and maintaining a constant force, ensuring the tip-sample contact during the scans. We use Nanonis real-time controllers for force feedback of the piezoresistive AFM cantilevers. For all measurement, a constant force of approximately 80 nN is applied which is suitable for achieving a nanoscale spatial resolution and a low z noise, without damaging the hBN surface (see Supplementary Fig. 2).
%~\ref{figs:AFM_images} 
\justify

\textbf{Device fabrication} \\
We use a dry transfer fabrication method with a polycarbonate (PC) stamp to fabricate the gated, hBN-encapsulated transition metal dichalcogenide devices. WSe$_2$ and MoSe$_2$ monolayers, few-layer graphene flakes for gates and contact and hBN for encapsulation are mechanically exfoliated from bulk crystals onto 300 nm SiO$_2$ on Si substrates.  The thickness of the hBN flakes, estimated from optical contrast, ranges from 15 to 30 nm. Firstly, we use a PC stamp to pick up the bottom hBN and the bottom few-layer graphene, which is then deposited onto a sapphire substrate with EBL-made electrodes (3 nm Cr/100 nm Au) at 180 $^\circ$C. The bottom gate structure was cleaned using contact-mode AFM. The top hBN flake, WSe$_2$ monolayer, MoSe$_2$ monolayer, and the few-layer-graphene contact are picked up in that order with a PC stamp and then placed onto the back gate at 180 $^\circ$C. Polymer residual from this transfer was washed off with chloroform at room temperature. We align the WSe$_2$ and MoSe$_2$ layers relying on their straight edges within 1$^\circ$ uncertainty (Supplementary Fig. 2).
%~\ref{figs:sample_opt}). 
Another round of AFM cleaning was performed on the entire stack before measurement.
\justify

\textbf{Optical measurements} \\
Photoluminescence spectroscopy, reflectivity, and emission lifetime measurements were performed at the base temperature of 4 K in an AttoDRY 1000 cryostat equiped with a 9T-superconducting magnet. The cryostat features a homemade insert including a confocal optical microscope, and two stacking of piezoelectric positioners and scanners (Attocube) to move the sample and the AFM tip independently. An achromatic objective (NA = $0.81$) focuses the laser beam to a spot size of $\sim$1 $\mu$m. 

To perform photoluminescence spectroscopy, we excite the sample using either a SpectraPhysics Matisse continuous-wave laser, with a central wavelength in the range $720-740$~nm, tuned to optimize the emission intensity at each sample spot, or a Picoquant LDH-IB-730-B operated in continous-wave mode. The excitation is linearly polarized. In both cases, the emitted light is collected through the same objective and sent in free space to a high-resolution spectrograph (Teledyne Princeton Instruments HSR-750, focal length 750mm). For photoluminescence measurements, we operate the spectrograph with a 1200 grooves/mm grating (blazed at 750 nm) and use a PyLoN charge coupled device (Teledyne Princeton Instruments). The spectrograph entrance slit is kept narrow to achieve a spectral resolution of approximately $50~\mu$eV % 3 pixels
at the emission energy. For reflectivity measurements, we excite the sample using a SuperK EVO EUL-10 laser. The reflected light is spread with a 300 grooves/mm grating (blazed at 750 nm), and recorded using a PIXIS-400 charge coupled device (Teledyne Princeton Instruments). 

For polarization-resolved PL measurement, polarization of the incident laser was controlled by using a polarizer together with a full-wave liquid crystal retarder. Circular-polarization-resolved measurements were performed with a $\lambda/4$ (achromatic, $690 - 1,200$~nm) wave plate placed before the Wollaston prism. The $\sigma$+ and $\sigma$− components of the polarization are collected simultaneously. Circular emission was converted into linearly polarized light after passing through the $\lambda/4$ plate. The Wollaston prism separates light into \textit{s}- and \textit{p}-components. Another $\lambda/4$ wave plate is placed after the Wollaston prism to convert the linearly polarized light into a circularly polarized signal. In all the magnetic field dependence measurements, the magnetic field is applied perpendicular to the plane of the sample.

To perform emission lifetime measurements, we employ a Picoquant LDH-IB-730-B in pulsed mode at a repetition rate of $1$~MHz, chosen to ensure the complete decay of the signal between consecutive laser pulses. The power is kept at $7$~nW, well above saturation of the two-level emitters. We couple the photoluminescence into a single-mode fiber ($780$-HP), and send it to an avalanche photo-diode (PerkinElmer SPCM-AQR-15-FC) connected to a photon counter (Picoquant PicoHarp 330). 
To measure the overall population decay of multiple quantum emitters, we select a spectral region of a few tens of nm using a combination of long- and short-pass filters placed before the fiber coupling. In contrast, to measure the decay of a single quantum dot, the signal is spectrally dispersed using a high-resolution, high-efficiency transmission grating (Ibsen Photonics, Raman FSTG-NIR1500-908), such that only a spectral region narrower than $0.5$~nm is coupled into the fiber and sent to the photodiode. %The acquisition times vary between $3000$ and $3600$~s.

%The pulsed experiments allow for a targeted initialization of a particular exciton population to the system, while the CW excitation establishes a steady-state excitonic density.  

%The excitation and emission arms also each featured a liquid crystal retarder (Thorlabs LCC1513-B), which enabled us to excite the sample (and collect emission of light) with well-defined linear or circular polarization.

We apply independent voltages to the graphene back gate and to the AFM conducting tip using two source meters (Keithley 2400), allowing us to tune the charge density and apply a displacement field to the sample. For device B (see Fig.~\ref{fig:doping}), the contact to the heterobilayer is grounded to a common ground shared by the source meters. For all other devices, the sample contact is left floating.

\justify
%\textbf{Emission Lifetime and Energy Simulation} \\
\textbf{Electric field simulations for tip-induced Stark-shift} \\
The Stark shifts were calculated from finite-element simulations of the in-plane and out-of-plane components of the electric field, performed using the Electrostatics module in COMSOL Multiphysics (see Supplementary Fig. 5).
%~\ref{figs:comsol}). 
We consider a two-dimensional axisymmetric geometry, with the symmetry axis coinciding with the tip axis. The simulated structure consists of a $40$-nm-thick hBN layer in contact with the tip, which is surrounded by vacuum. The tip has a radius of curvature $R_t = 22$~nm, and a truncated pyramid body with top base of $2R_t\cos(\pi/4)$ and bottom base of $1.49~h$, where $h_t=800$~nm is the tip height. The height-to-base ratio and the radius $R_t$ correspond to the experimental values extracted from SEM imaging (see Supplementary Fig. 1).
%~\ref{figs:setup}). 
The bottom boundary of the hBN layer and the tip are held at constant potentials of $V_\text{bg}=0$~V and $V_\text{tip}=-1$~V, respectively. 
The electric field is evaluated on a plane passing through the center of the hBN layer. The tip height $h_t$ and the dimensions of the simulation domain are chosen based on convergence tests to ensure that the calculated electric field is not affected by the finite boundaries.

\justify
\textbf{Lifetime simulations} \\
The tip-induced modification of the LPDOS was numerically simulated, as shown in Supplementary Fig. 13.
%~\ref{figs:lifeVSsim}. The computation is based on the calculation of the dyadic Green's function for an homogeneous medium ${\bf G}(r,k)=e^{ikr}/4\pi r (c_1 \mathbb{1} + c_2 \hat{n} \otimes \hat{n})$, where $c_1 = 1 + i/kr - 1/kr^2$ and $c_2 = -1 - 3i/kr + 3/kr^2$. This allows us to calculate the electric field induced by a radiating dipole ${\bf p}$ at a distance $r=|{\bf r}|$ from the dipole, i.e. ${\bf E}(r,k) = (k^2/\epsilon_0)~{\bf G}(r,k) {\bf p}$.
We consider an in-plane dipole radiating at $\lambda_0=900$~nm ($k_0=2\pi/\lambda_0$), surrounded by hBN, and placed at $\Delta z =20$~nm from a vacuum-hBN interface. We use $\tilde{\epsilon}=4.5$ as average dielectric constant for hBN, and $\epsilon_v=1$ for vacuum. 
We first calculate the decay-rate contribution of the vacuum-hBN interface $\gamma_\text{int} = R\frac{6\pi\epsilon_0}{\tilde{k}^3}|{\bf p}|^2 \Imag({\bf p} \cdot {\bf E}(r_\text{image},\tilde{k}))$ where $\tilde{k}=k_0\sqrt{\tilde{\epsilon}}$ is the wavevector in the hBN layer, $r_\text{image}$ is the distance of the image dipole from the source, and $R =(\tilde{\epsilon}-\epsilon_v)/(\tilde{\epsilon}+\epsilon_v)$ is the Fresnel reflection coefficient.
%by computing the electric field induced by the image dipole at the emitter position. 
This contribution is independent on the tip position. 
Then, at each tip-dipole lateral displacement $\Delta x$, we calculate the electric field ${\bf E}(r_\text{tip}, \tilde{k})$ at the tip position, where $r_\text{tip}=\sqrt{\Delta z^2+ \Delta x^2}$. 
After transmission through the hBN-vacuum interface, this electric field induces a dipole moment ${\bf p}_\text{tip} = T_1 \alpha_\text{tip}{\bf E}(r_\text{tip},\tilde{k})$ in the metallic tip, where $T_1=2\tilde{\epsilon}/(\tilde{\epsilon}+\epsilon_v)$ is the Fresnel transmission coefficient from hBN to vacuum ($T_2=\epsilon_vT_1/\tilde{\epsilon}$ from vacuum to hBN).
We model the metallic tip as a polarizable sphere of radiatively-corrected polarizability $\alpha_\text{tip} = \alpha_0/(1-i\alpha_0k_0^3/6\pi\epsilon_0)$, where $\alpha_0 = 4\pi\epsilon_0R_t^3 (\tilde{\epsilon}-\epsilon_v)/(\tilde{\epsilon}+\epsilon_v)$. 
The tip radius $R_t=50$~nm is fixed to the value measured from the SEM image of the tip employed in the data of Fig.~\ref{fig:lifetime}. 
The tip dipole moment generates an electric field ${\bf E}_\text{tip}(r_\text{tip},k_0) = (T_2 k_0^2/\epsilon_0)~{\bf G}(r_\text{tip},k_0) {\bf p}_\text{tip}$ at the position of the dipolar emitter, which contributes to the decay rate as $\gamma_\text{tip} \equiv \gamma_\text{tip}(\Delta x) = \frac{6\pi\epsilon_0}{\tilde{k}^3}|{\bf p}|^2 \Imag({\bf p_\text{tip}} \cdot {\bf E}_\text{tip}(r_\text{tip},k_0))$. 
Finally, we calculate the lifetime modification $\tau/\tau_\text{far}$, where $\tau(\Delta x)^{-1} = \Gamma(\Delta x) = 1 +\gamma_\text{int}+\gamma_\text{tip}(\Delta x)$ and $\tau_\text{far} \equiv \tau(\Delta x \rightarrow \infty)$.
The results are in good agreement with those obtained using more sophisticated semi-analytical field-expansion methods based on Green's functions~\cite{novotny_single_1996}.

%TC:endignore
% ^ this is for the word counter

% Methods-only references
%\justify
%\textbf{Methods-only references}

% Figures/Figure Captions

\end{document}

%Title
\centering{\bf Supplementary Information}
\title{Super-resolution Control of Individual Two-dimensional Quantum Emitters}

%Author List/Affiliations
\author{Bosai Lyu$^{\dagger}$}
\affiliation{Department of Quantum Matter Physics, University of Geneva, Geneva, Switzerland}
%\affiliation{Department of Physics, Emory University, 30322 Atlanta, Georgia, USA}

\author{Valeria Vento$^{\dagger}$}
\affiliation{Department of Quantum Matter Physics, University of Geneva, Geneva, Switzerland}

\author{Ludivine Fausten}
\affiliation{Department of Quantum Matter Physics, University of Geneva, Geneva, Switzerland}

\author{Daniel Suárez-Forero}
\affiliation{Department of Quantum Matter Physics, University of Geneva, Geneva, Switzerland}

\author{Klevis Domi}
\affiliation{Department of Quantum Matter Physics, University of Geneva, Geneva, Switzerland}

\author{Kenji Watanabe}
\affiliation{Research Center for Electronic and Optical Materials, National Institute for Materials Science, 1-1 Namiki, Tsukuba 305-0044, Japan}

\author{Takashi Taniguchi}
\affiliation{International Center for Materials Nanoarchitectonics, National Institute for Materials Science,  1-1 Namiki, Tsukuba 305-0044, Japan}

\author{Alberto Morpurgo}
\affiliation{Department of Quantum Matter Physics, University of Geneva, Geneva, Switzerland}
\affiliation{Geneva Quantum Centre, University of Geneva, Geneva, Switzerland}

\author{Iaroslav Gaponenko}
\affiliation{Department of Quantum Matter Physics, University of Geneva, Geneva, Switzerland}

\author{Patrycja Paruch}
\affiliation{Department of Quantum Matter Physics, University of Geneva, Geneva, Switzerland}
\affiliation{Geneva Quantum Centre, University of Geneva, Geneva, Switzerland}

\author{Ajit Srivastava$^{\ast}$}
\affiliation{Department of Quantum Matter Physics, University of Geneva, Geneva, Switzerland}
\affiliation{Geneva Quantum Centre, University of Geneva, Geneva, Switzerland}

\maketitle
\justify
$^{\dagger}$These authors contributed equally to this work \\
$^{\ast}$Correspondence to: ajit.srivastava@unige.ch \\

% Abstract:

%TC:ignore
% ^ this is for the word counter

%To perform the DCP measurements, we either a polarizer or with a full-wave liquid-crystal retarder 
\justify
\textbf{Contents}

\begin{figure}
\includegraphics[width = 6in]{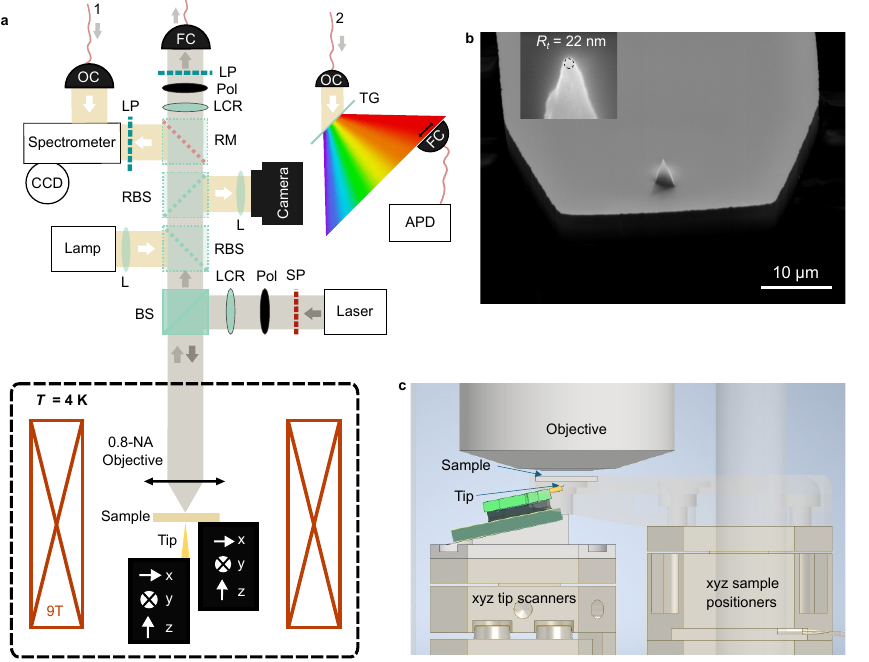}
\caption{{\bf Experimental setup.} {\bf a,} Sketch of the QOSMic. Acronym legend: LP - long-pass filter; Pol - linear polarizer; LCR - liquid crystal retarder; L - lens; BS - beamsplitter; RBS - removable beamsplitter; RM - removable mirror; FC - fiber coupler; OC - output coupler; TG - transmission grating; APD - avalanche photodiode;
{\bf b,} Scanning electron microscope (SEM) image of the self-sensing cantilever. Inset: High-resolution SEM image of a new conductive tip showing a curvature radius of $\sim$22 nm. {\bf c,} Schematic of the setup close to the sample.}
\label{figs:setup}
\end{figure}

\noindent
\textbf{Supplementary Fig. ~\ref{figs:setup}.} Experimental setup.

\begin{figure}
\includegraphics[width = \textwidth]{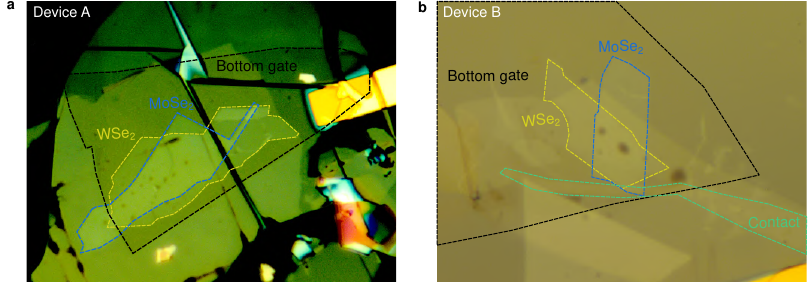}
\caption{{\bf Optical microscope images of the measured devices. } {\bf a,} Device A. 
{\bf b,} Device B. }
\label{figs:sample_opt}
\end{figure}

\noindent
\textbf{Supplementary Fig. ~\ref{figs:sample_opt}.} Optical microscope image of devices.

\begin{figure}
\includegraphics[width = \textwidth]{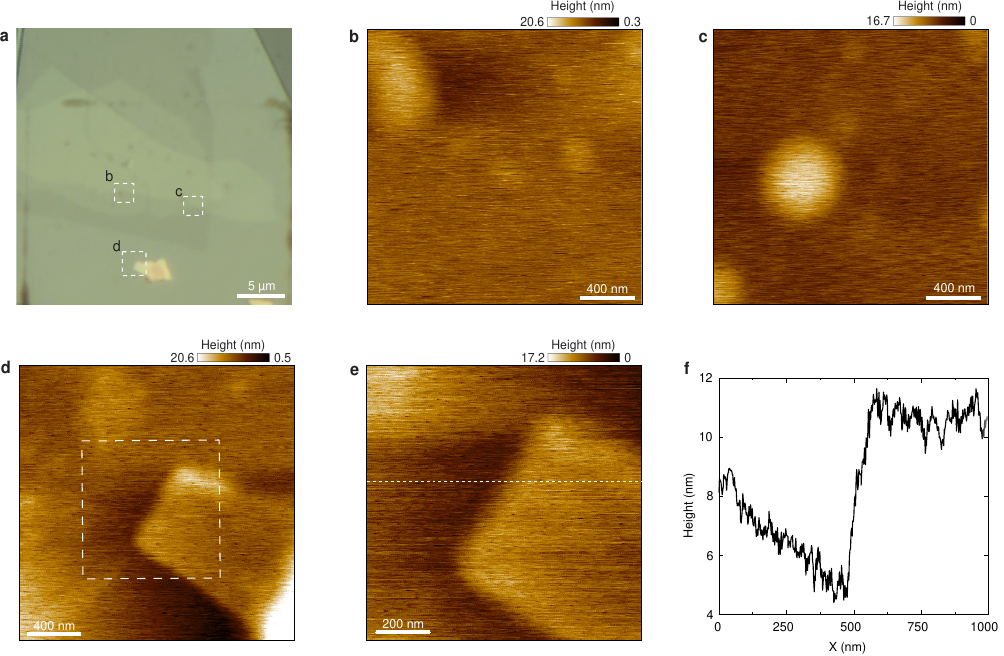}
\caption{{\bf AFM topographic images measured at 4 K. } {\bf a,} Optical microscope image of device A. 
{\bf b} and {\bf c,} AFM images of spot II and spot IV including their surrounding regions. {\bf d} Scan of an hBN structure. {\bf e,} Zoom-in scan in the dashed region of {\bf d}. {\bf f,} Line profile along the dashed line in panel {\bf e}, showing a peak-to-peak height noise below 2 nm.}
\label{figs:AFM_images}
\end{figure}

\noindent
\textbf{Supplementary Fig. ~\ref{figs:AFM_images}.} AFM topographic images taken at 4 K.

\begin{figure}
\includegraphics[width = \textwidth]{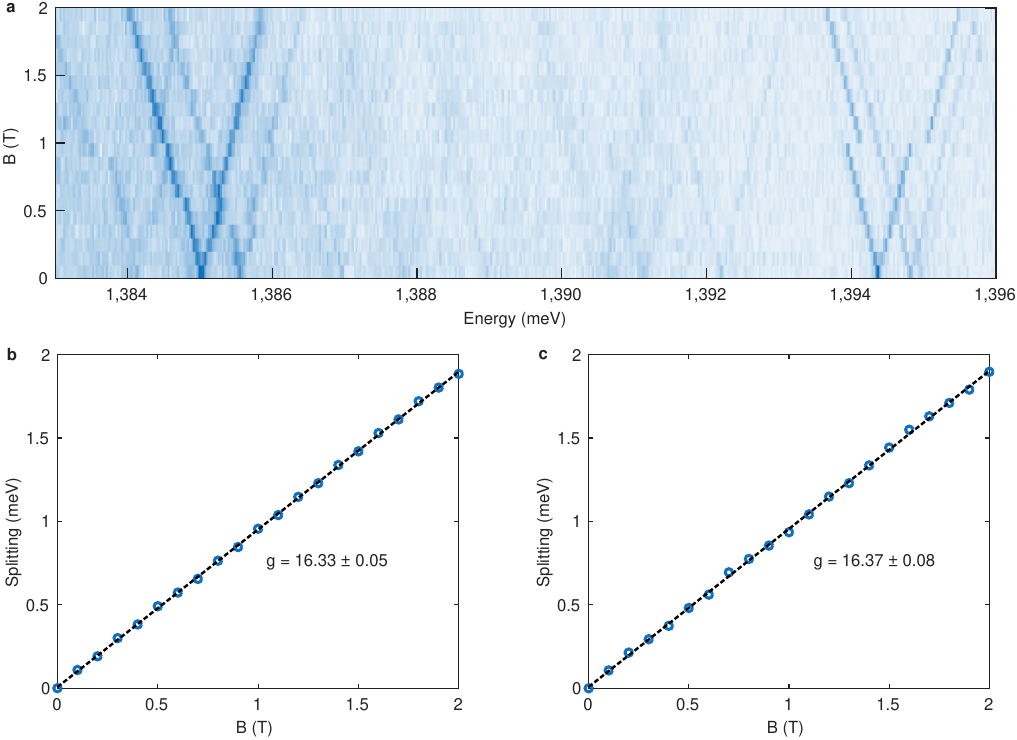}
\caption{{\bf \textit{g}-factor. } {\bf a,} PL emission as a function of the out-of-plane magnetic field \textit{B} in spot IV of device A. 
{\bf b} and {\bf c,} Splitting energies of dots at 1,385 meV and 1,394.4 meV as a function of \textit{B}. They have the same \textit{g}-factor of about 16, consistent with H-stacking registry.}
\label{figs:g-factor}
\end{figure}

\noindent
\textbf{Supplementary Fig. ~\ref{figs:g-factor}.} \textit{g}-factor.

\begin{figure}
\includegraphics[width = \textwidth]{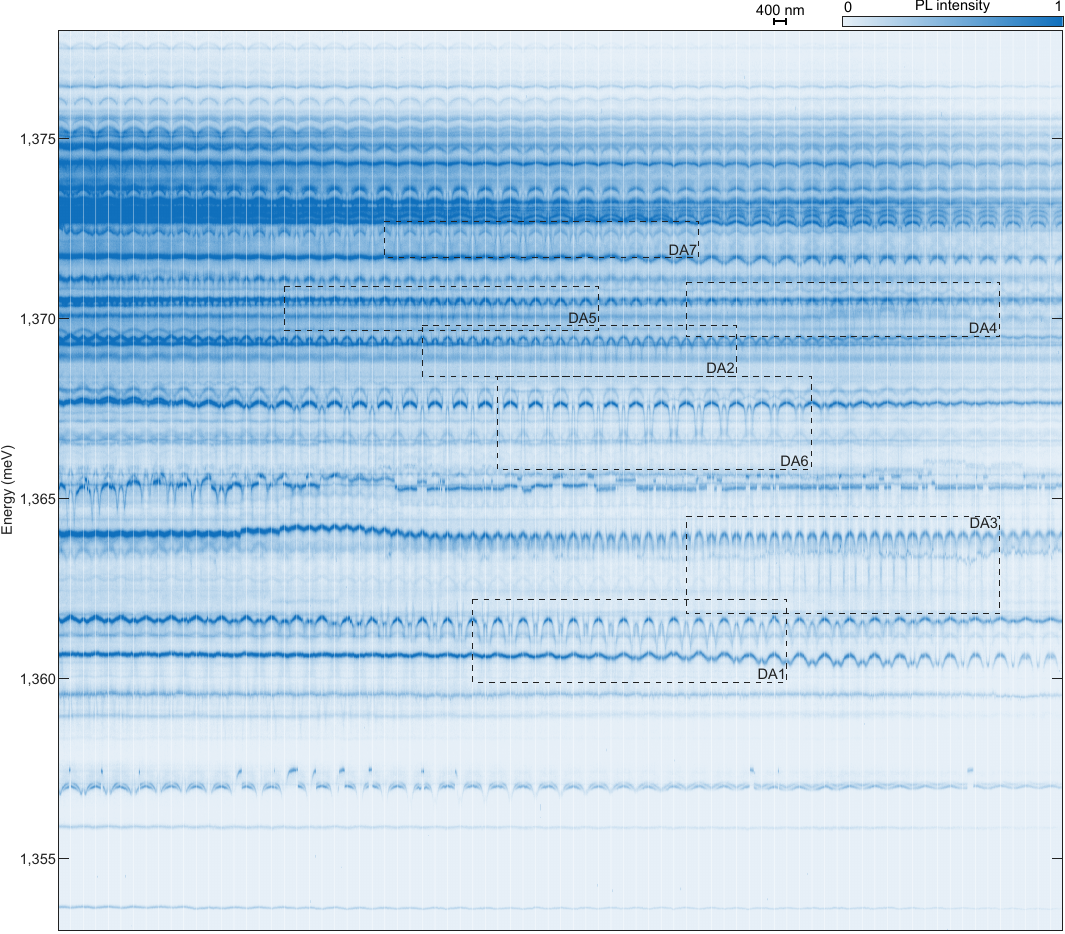}
\caption{{\bf Tip-position dependence of the PL emission spectrum from spot III of device A.} The spectra are collected sequentially while the tip is raster scanned on a $400$-nm square, with $5$-nm step size and $V_\text{tip} = -1.2$~V. White lines separate the tip line scans along the $y$ direction (fast axis) at a fixed $x$ coordinate (slow axis). The dashed rectangles indicates the dots DA1-7, as in Fig.~\ref{fig:superres}f and Supplementary Fig.~\ref{figs:superres_zoom}.}
\label{figs:superres}
\end{figure}

\noindent
\textbf{Supplementary Fig. ~\ref{figs:superres}.} Tip-position dependence of the PL emission from spot III of device A.

\noindent
\textbf{Supplementary Fig. ~\ref{figs:superres_zoom}.} Zoom-in view of dots in Supplementary Fig. ~\ref{figs:superres}.

\begin{figure}
\includegraphics[width =\textwidth]{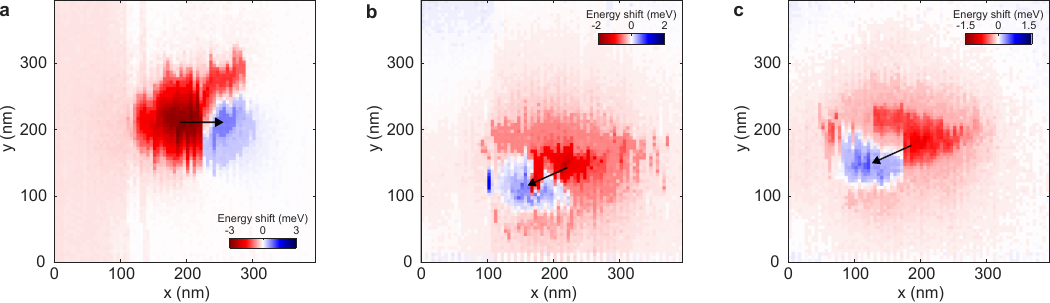} 
\caption{ {\bf In-plane dipole orientation.} Dots in spot II ({\bf a}) and spot III ({\bf b-c}) of device B possess different in-plane dipole directions. Panels {\bf b} and {\bf c} show dot DA1 and DA2 of Fig.~\ref{fig:superres}f respectively. }
\label{figs:dipole_orientation}
\end{figure}

\noindent
\textbf{Supplementary Fig. ~\ref{figs:dipole_orientation}.} In-plane dipole orientation.

\noindent
\textbf{Supplementary Note 1.} Spatial resolution.

\begin{figure}
\includegraphics[width = \textwidth]{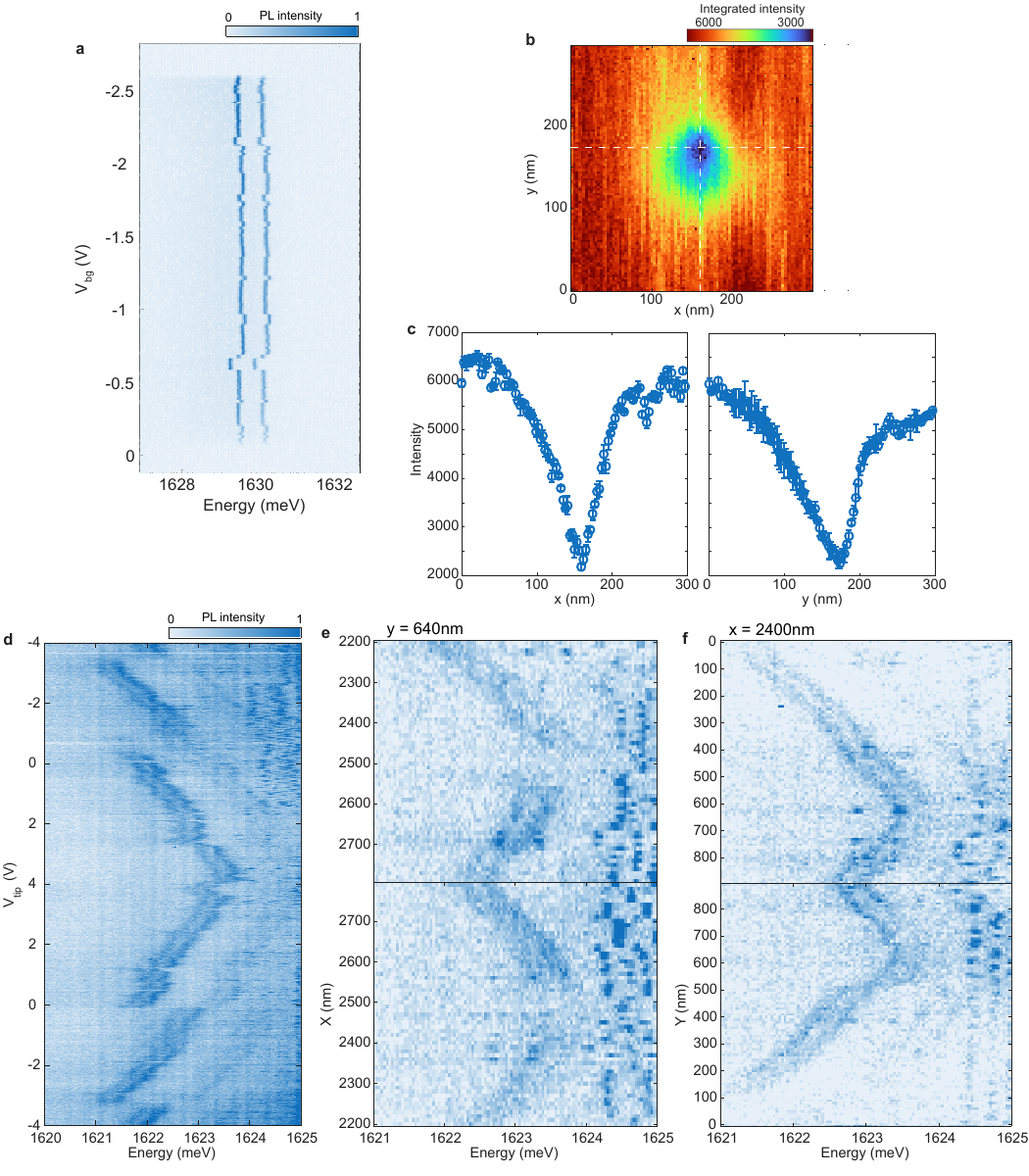}
\caption {{\bf Localizations of quantum emitters in WSe$_2$ monolayers.} {\bf a,} $V_\text{bg}$ dependence of the PL emission of a quantum emitter in spot II of device B. {\bf b,} Tip-position dependence of the integrated PL intensity, with a step size of 3 nm and $V_\text{tip} = V_\text{bg} = 0$~V. {\bf c,} Line profiles taken along the $x$ and $y$ dashed lines in panel {\bf b}. Each blue circle and corresponding error bar are the averaged value and standard deviation of three neighboring lines. {\bf d,} Tip-voltage-dependent PL emissions of a quantum emitter in spot V of device A (tip-dot distance approximately 50 nm). The linear-like Stark shift implies this emitter could possess a permanent dipole moment. {\bf b} and {\bf c,} $x$ and $y$ line scans measured at $V_\text{tip} = 4$~V, $V_\text{bg} = 0$~V. } 
\label{figs:monolayerdot}
\end{figure}

\noindent
\textbf{Supplementary Fig. ~\ref{figs:monolayerdot}.} Localizations of quantum emitters in WSe$_2$ monolayers.

\noindent
\textbf{Supplementary Note 2.} Localizations of quantum emitters in WSe$_2$ monolayers.

\begin{figure}
\includegraphics[width = \textwidth]{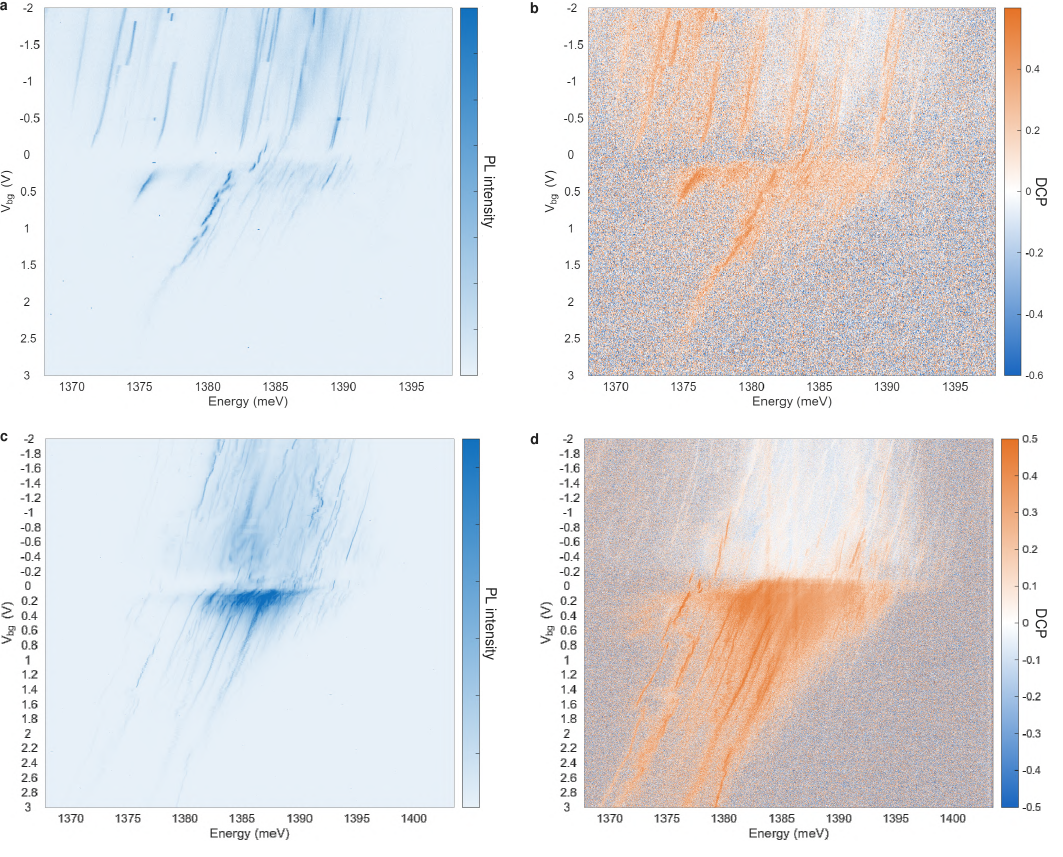}
\caption{{\bf Bottom-gate voltage dependence of PL intensity and DCP. } {\bf a,} {\bf b,} PL intensity and DCP as a function of $V_\text{bg}$ measured at spot I of device B.
{\bf c,} {\bf d,} PL intensity and DCP as a function of $V_\text{bg}$ measured at spot II of device B. As expected, trions show a larger DCP compared to neutral excitons.}
\label{figs:DCP}
\end{figure}

\noindent
\textbf{Supplementary Fig. ~\ref{figs:DCP}.} Bottom gate voltage dependence of PL intensity and DCP.

\begin{figure}
\includegraphics[width = \textwidth]{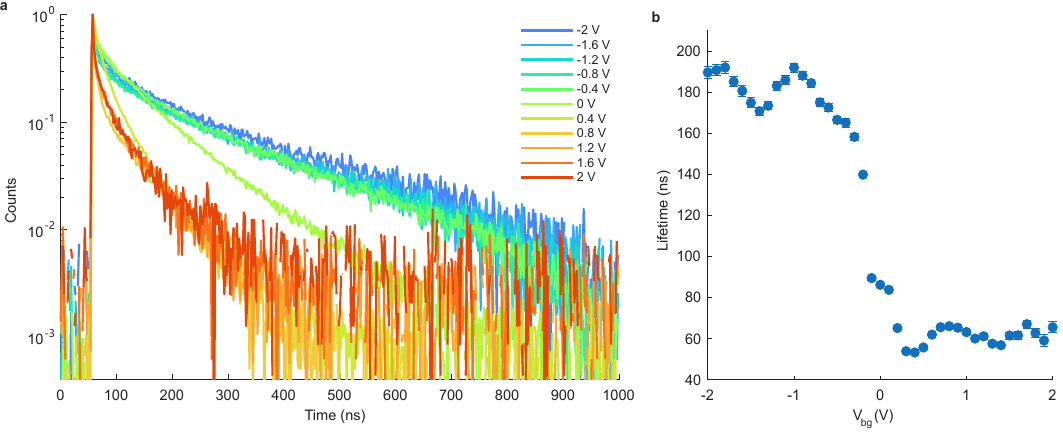}
\caption{{\bf Bottom-gate voltage dependence of PL lifetime. } {\bf a,} PL emission decay as a function of $V_\text{bg}$ with tip retracted.
{\bf b,} Lifetime extracted by exponential fit in a restricted time range as a function of $V_\text{bg}$. }
\label{figs:All_dots_lifetime}
\end{figure}

\noindent
\textbf{Supplementary Fig. ~\ref{figs:All_dots_lifetime}.} Bottom gate voltage dependence of PL lifetime.

\begin{figure}
\includegraphics[width = \textwidth]{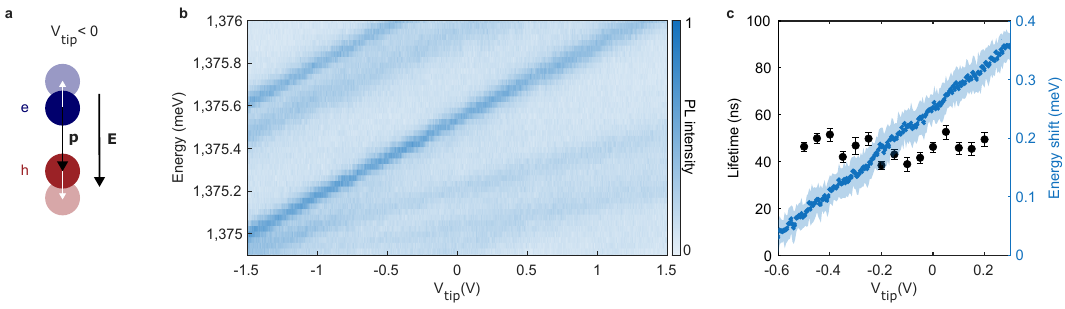}
\caption{{\bf Electric-field effect on the lifetime.} {\bf a,} Sketch of the electric field effect on the electron-hole displacement. For a negative tip bias, the electric field is parallel to the dipole moment, and therefore it pulls the electron and hole apart. {\bf b,} Tip-bias dependence of the PL energy of a dot.
{\bf c,} Lifetime of the same dot as a function of tip bias (black circles). The energy shift (blue circles) is extracted from panel {\bf b} by a Lorentzian fit. Shadow corresponds to the Lorentzian FWHM. We observe that the lifetime is constant while the dot undergoes a total energy shift $2$ times larger than in the data of Fig.~\ref{fig:lifetime}c. This further confirms that the electric field plays a negligible role in the lifetime modification presented in the main text.}
\label{figs:lifeVSshift}
\end{figure}

\noindent
\textbf{Supplementary Fig. ~\ref{figs:lifeVSshift}.} Electric-field effect on the lifetime.

\begin{figure}
\includegraphics[width = \textwidth]{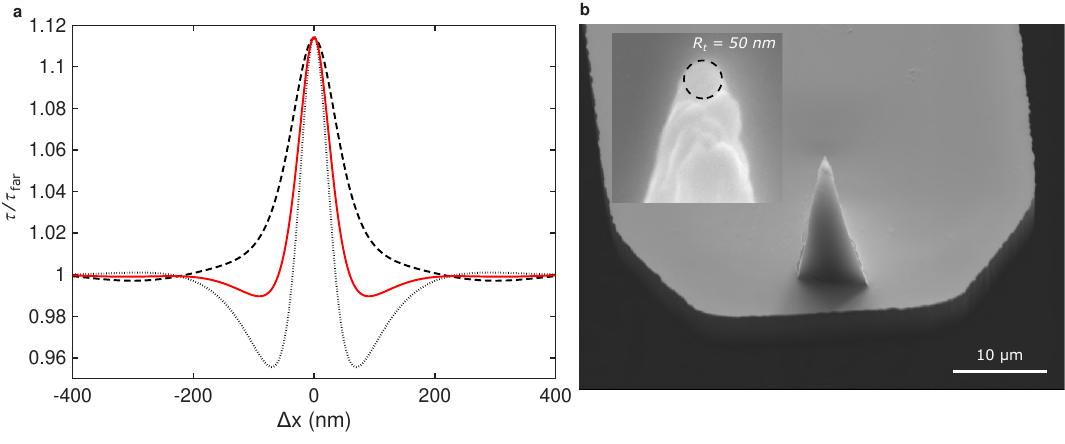}
\caption{{\bf Numerical simulation of tip-induced lifetime modification. }
{\bf a,} Normalized lifetime $\tau/\tau_\text{far}$ of a radiating dipole at $900$~nm as a function of the tip lateral displacement $\Delta x$ from the dipole; $\tau_\text{far}$ denotes the lifetime when the tip is far from the emitter. The black dashed line corresponds to an in-plane dipole along $x$ (direction of tip motion); the black dotted line corresponds to an in-plane dipole along $y$; the red line is the average of the two curves (same as in Fig.~\ref{fig:lifetime}). The curves are obtained by numerical calculation of the dyadic Green's function, as explained in Methods. The tip is modeled as a polarizable sphere of radius $R_t=50$~nm, which was measured by SEM imaging in panel {\bf b}. {\bf b} SEM image of the tip used for the data in Fig.~\ref{fig:lifetime}.}
\label{figs:lifeVSsim}
\end{figure}

\noindent
\textbf{Supplementary Fig. ~\ref{figs:lifeVSsim}.} Simulated tip-induced lifetime modification

\clearpage

\clearpage

\clearpage

\begin{figure}
\includegraphics[width = \textwidth]{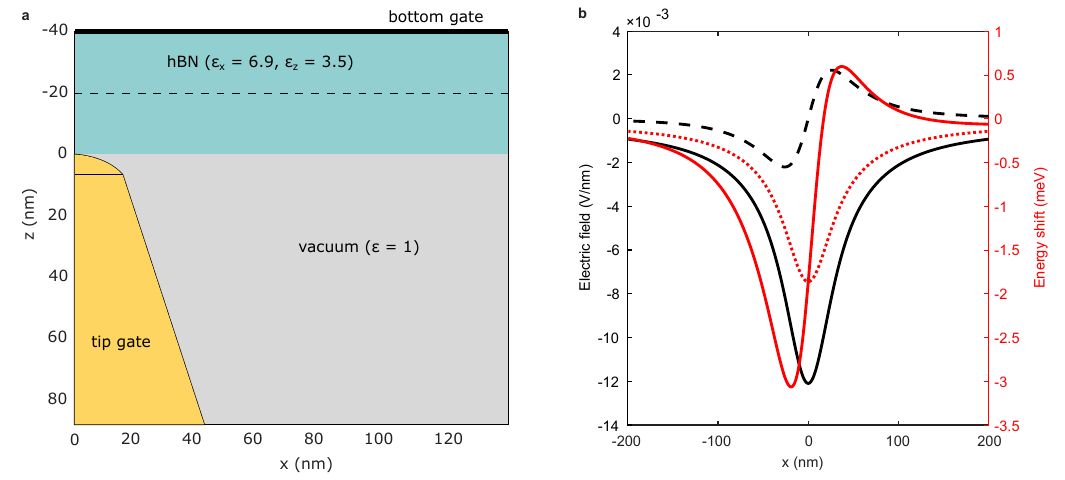}
\caption{{\bf Finite-element simulations of the electric field due to V$_\mathrm{tip}$.} {\bf a,} Sketch of the COMSOL 2D geometry (zoom-in) with axial symmetry around the tip axis $x=0$, as described in Methods. The tip height is $h_t=800$~nm. The height-to-base ratio of $1.49$ and the tip curvature radius $R_t = 22$~nm are extracted from SEM imaging (see Supplementary Fig.~\ref{figs:setup}). The dielectric constants of the relevant materials are indicated in the figure.
{\bf b,} Computed electric field components $E_x$ (black dashed) and $E_z$ (black solid) along a line passing through the TMD plane $z=-20$~nm (dashed in panel {\bf a}). The tip- and bottom-gate voltages are $V_\text{bg}=0$~V and $V_\text{tip}=-1$~V respectively. The red line (same as in Fig.~\ref{fig:superres}c) is the calculated energy shift for a dipole with electron-hole displacement ${\bf d} = (3.5,~ 0,~ 0.7)$~nm, as explained in the main text. In this case, the $x$ axis corresponds to the displacement of the dipole with respect to the tip. We remark that the maximum red shift does not occur when the dot is aligned with the tip ($x=0$), but rather at a lateral displacement of approximately $19$~nm along the in-plane dipole component. The red dotted line is calculated energy shift for a dipole oriented out-of-plane, i.e. ${\bf d} = (0,~ 0,~ 0.7)$~nm. }
\label{figs:comsol}
\end{figure}

\begin{figure}
\includegraphics[width = \textwidth]{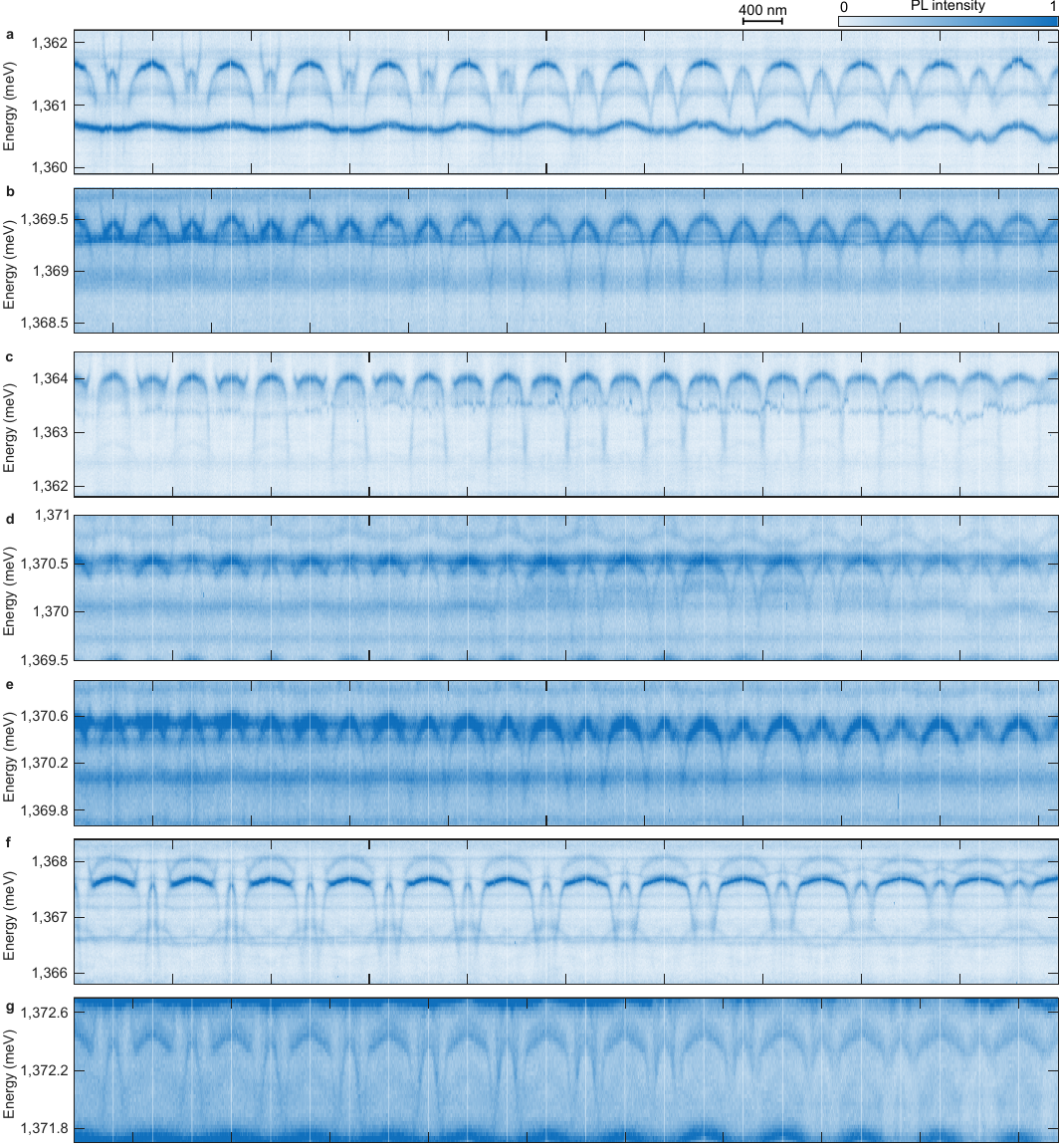}
\caption{{\bf Zoom-in view of dots in Supplementary Fig.~\ref{figs:superres}}. {\bf a}–{\bf g} correspond to the regions marked by dashed rectangles (dots DA1–7) in Supplementary Fig.~\ref{figs:superres}. } 
\label{figs:superres_zoom}
\end{figure}

\ref{fig:doping}

\clearpage
\section*{Supplementary Note 1: Spatial resolution}
The spatial resolution of QOSMic depends on our ability to distinguish two emitters by their relative Stark shift. In order to detect the Stark shift, we used a spectrometer-CCD based detection scheme (see Methods) but in principle other schemes employing single-photon detectors and higher resolution gratings can give higher spectral resolution. To estimate the spatial resolution based on our current detection scheme, we can consider two emitters with the same emission energy. The electric field from the tip should then be able to cause a relative Stark shift larger than one CCD pixel width ($\sim$50 $\mu$eV). Although the Stark shift can be made larger by increasing V$_\mathrm{tip}$, it cannot be made arbitrarily large as this can cause quenching of emission and other unwanted effects. Hence, we can choose a conservative value of V$_\mathrm{tip}$ = -1 V. In addition, we assume a 20 nm thick hBN as a spacer distance which sets the minimum distance between the tip and the emitter. Under these assumptions, the spatial resolution is determined jointly by the spatial gradient of the tip-induced Stark shift and the spectroscopic precision with which the emitter transition can be localized. 

\begin{equation}
\delta x_{\min}
=
\frac{\delta \mathcal{E}_{\min}}
{\left|\dfrac{\partial \mathcal{E}}{\partial x}\right|}.
\end{equation}

The tip-induced energy shift profile for an out-of-plane dipole (see the red dotted curve in Supplementary Fig.~\ref{figs:comsol}b) exhibits a maximum spatial gradient of approximately $63 ~\mu$eV/nm, at a tip bias of -1 V. This would imply a resolution of about a nm which is unrealistic as the spatial extent of quantum emitters is expected to be a few nm. Thus, we can conclude that the technique is in principle not limited by the ability to detect Stark-shifts but practically limited by the minimum separation between emitters which might be determined by the moir\'e length $\sim$ 10 nm. 

%only limited by the minimum distance between dots in a sample. 
%If we consider the minimum possible spacing of 6 nm
%Let's consider two emitters with the same emission energy and resolution-limited linewidth. The spatial resolution is determined jointly by the spatial gradient of the tip-induced Stark shift and the spectroscopic precision with which the emitter transition can be localized. 

%\begin{equation}
%\delta x_{\min}
%=
%\frac{\delta E_{\min}}
%{\left|\dfrac{\partial E}{\partial x}\right|}.
%\end{equation}

%Let's take the MoSe2/WSe2 heterobilayer sample hosting emitter only possess a out-of-plane permenent dipole moment as a example. The device has top- and bottom-hBN of 20-nm-thickness. The tip-induced energy shift profile (Fig.~\ref{figs:comsol}) exhibits a maximum spatial gradient of approximately $60 ~\mu$eV/nm, at a tip bias of -1 V. The minimal spectroscopic resolution is 50 $50~\mu$eV (one pixel). If we place the tip approximately 25 nm to the emitter where the spatial energy shift gradient is maxmal, a resolution of 1 nm can be expected.

Remarkably, the spatial resolution gets improved if the dot has an additional in-plane dipole moment, as we experimentally observe in Fig.~\ref{fig:superres}. In this case, the tip-induced Stark shift (solid red curve in Supplementary Fig.~\ref{figs:comsol}b) shows a maximum slope of approximately $124 ~\mu$eV/nm, which is two-times larger than the one for a purely out-of-plane dipole.

 %Since the minimum spectral resolution we can achieve with the current spectrograph is of $20~\mu$eV, we can in principle resolve two dots at a distance below $1$~nm. We remark that the presence of an in-plane dipole gives a further advantage in the resolution, since it produces a dispersive line shape in the energy shift which results in a higher slope. Moreover, this slope can be further increased by increasing the tip bias, which is here kept at $-1$~V, well below any damage threshold.

\clearpage
\section*{Supplementary Note 2: Localizations of quantum emitters in WSe$_2$ monolayers}
QOSMic allows for the localization of dots in monolayer WSe$_2$ by the tip-induced PL intensity change and Stark shift in emission energy, in addition to the interlayer dots in MoSe$_2$/WSe$_2$ heterobilayers discussed in the main text. Because of the intralayer nature, the majority of monolayer dots lacks a permanent dipole moment, such that no Stark shift is expected. Indeed, for example, in spot III of device B, we observe a neutral doublet with a splitting of approximately 0.6 meV (Supplementary Fig.~\ref{figs:monolayerdot}a). This dot does not respond to the electric field applied by the bottom gate and is doped when $V_\text{bg}$ is large enough. By bringing the tip close to the dot without applied gate voltages, we observe a significant decrease in the PL intensity (Supplementary Fig.~\ref{figs:monolayerdot}b). The line profile of the PL intensity reveals a resolution of approximately 50 nm (Supplementary Fig. ~\ref{figs:monolayerdot}c). This decrease in intensity could be attributed to the suppression of the local photoinc density of states, as discussed in the ``Control of radiative dynamics" section in the main text.

In spot V of device A, we find a dot showing a splitting of approximately 0.35 meV, which unexpectedly exhibits a linear-like blue-shift as increasing $V_\text{tip}$ (Supplementary Fig.~\ref{figs:monolayerdot}d). With $V_\text{tip} = 4$ V, we conduct line scans across the dot as shown in (Supplementary Fig.~\ref{figs:monolayerdot}e and ~\ref{figs:monolayerdot}f). A blue shift in the emission energy is always observed as bringing the tip closer to the dot, which implies that this dot possibly possess an out-of-plane permanent dipole moment. Also, when the tip is moving precisely across the dot (Supplementary Fig.~\ref{figs:monolayerdot}e), the PL intensity decreases, consistent with the behavior of the other dot described above.
\justify
\clearpage

%\clearpage

%\bibliographystyle{apsrev}
%\bibliography{references}